\documentclass[a4paper,11pt]{article}
\usepackage{jinstpub} 
\usepackage{lineno}
\usepackage{orcidlink}
\usepackage{comment}
\usepackage{caption}
\usepackage{subcaption}
\usepackage{upgreek}
\usepackage{float}
\usepackage{ragged2e}
\usepackage{pdfpages}
\usepackage{bbold}
\usepackage{url}
\usepackage[utf8]{inputenc}
\usepackage{footmisc}
\usepackage{siunitx}
\usepackage{jinstpub}
\title{Simulation and optimization of the Active Magnetic Shield of the n2EDM experiment}

\author[a]{N.~J.~Ayres\orcidlink{0000-0002-5565-0807}}
\author[b]{G.~Ban}
\author[c]{G.~Bison\orcidlink{0000-0002-0624-6672}}
\author[d]{K.~Bodek}
\author[a]{V.~Bondar}
\author[e]{T.~Bouillaud}
\author[a,c]{G.~L.~Caratsch}
\author[f,1]{E.~Chanel}
\author[a,c]{W.~Chen\orcidlink{0009-0008-1977-6833}}
\author[g]{C.~Crawford\orcidlink{0000-0002-1932-4334}}
\author[e]{V.~Czamler\orcidlink{0009-0005-3192-9984}}
\author[a,c]{C.~B.~Doorenbos}
\author[a]{S.~Emmeneger}
\author[a,*]{S.~K.~Ermakov\note[*]{Corresponding Authors}\orcidlink{0009-0005-0518-0703}}

\author[e]{M.~Ferry\note[1]{Present address: Institut Laue–Langevin, CS 20156, 38042 Grenoble Cedex 9, France}}
\author[h]{M.~Fertl\orcidlink{0000-0002-1925-2553}}
\author[f]{A.~Fratangelo}
\author[b]{D.~Galbinski\orcidlink{0009-0002-0214-2064}}
\author[i]{W.~C.~Griffith}
\author[j]{Z.~D.~Grujic\orcidlink{0000-0003-0802-5782}}
\author[a,c,*]{K.~Kirch\orcidlink{0000-0002-1720-7636}}
\author[a,c,2]{V.~Kletzl\note[2]{Present address: Marietta Blau Institute for Particle Physics, 1010 Vienna, Austria}}
\author[a]{J.~Krempel}
\author[c]{B.~Lauss\orcidlink{0000-0002-1986-391X}}
\author[b]{T.~Lefort}
\author[b]{A.~Lejuez}
\author[e]{K.~Michielsen}
\author[f]{J.~Micko}
\author[a,3]{P.~Mullan\orcidlink{0009-0002-8926-010X}\note[3]{Present address: Albert-Ludwigs-Universität Freiburg, Physikalisches Institut, 79104 Freiburg, Germany}}
\author[b]{O.~Naviliat-Cuncic\orcidlink{0000-0001-5082-2131}}
\author[f]{F.~M.~Piegsa\orcidlink{0000-0002-4393-1054}}
\author[e]{G.~Pignol\orcidlink{0000-0001-7086-0100}}
\author[f]{C.~Pistillo}
\author[c]{I.~Rienäcker\orcidlink{0000-0002-7411-2481}}
\author[c]{D.~Ries}
\author[e]{S.~Roccia\orcidlink{0009-0004-4752-5442}}
\author[d]{D.~Rozpędzik}
\author[a]{L.~Sanchez-Real~Zielniewicz}
\author[a,c]{N.~von~Schickh}
\author[c]{P.~Schmidt-Wellenburg\orcidlink{0000-0001-5474-672X}}
\author[a,c]{E.~P.~Segarra\orcidlink{0000-0001-5606-0350}}
\author[a]{L.~Segner}
\author[k]{N.~Severijns\orcidlink{0000-0003-4792-9362}}
\author[e,4]{K.~Svirina\orcidlink{0000-0002-2679-9020}\note[4]{Present address: Institut Laue–Langevin, CS 20156, 38042 Grenoble Cedex 9, France and Physikalisches Institut, Universitat Heidelberg, Im Neuenheimer Feld 226, 69120 Heidelberg, Germany}}
\author[f]{J.~Thorne\orcidlink{0000-0002-3905-5549}}
\author[k]{J.~Vankeirsbilck}
\author[c,5]{N.~Yazdandoost\note[5]{Present address: TRIUMF, Vancouver, British Columbia V6T 2A3, Canada}\orcidlink{0000-0001-6768-795X}}
\author[d]{J.~Zejma\orcidlink{0000-0003-4843-1451}}
\author[a,*]{N.~Ziehl\orcidlink{0000-0002-0136-3300}}
\author[c]{G.~Zsigmond\orcidlink{0000-0003-3476-751X}}

\affiliation[a]{ETH Zürich, Institute for Particle Physics and Astrophysics, CH-8093 Zürich, Switzerland}
\affiliation[b]{Normandie Univ, ENSICAEN, UNICAEN, CNRS/IN2P3, LPC Caen, 14000 Caen, France}
\affiliation[c]{Paul Scherrer Institut, CH-5232 Villigen PSI, Switzerland}
\affiliation[d]{Marian Smoluchowski Institute of Physics, Jagiellonian University, 30-348 Cracow, Poland}
\affiliation[e]{LPSC, Université Grenoble Alpes, CNRS/IN2P3, Grenoble, France}
\affiliation[f]{Laboratory for High Energy Physics and Albert Einstein Center for Fundamental Physics, University of Bern, CH-3012 Bern, Switzerland}
\affiliation[g]{University of Kentucky, Lexington, USA}
\affiliation[h]{Institute of Physics, Johannes Gutenberg University Mainz, D-55128 Mainz, Germany}
\affiliation[i]{University of Sussex, Department of Physics and Astronomy, Falmer, Brighton BN1 9QH, UK}
\affiliation[j]{Institute of Physics Belgrade, University of Belgrade, 11080 Belgrade, Serbia}
\affiliation[k]{Institute for Nuclear and Radiation Physics, KU Leuven, B-3001 Leuven, Belgium}

\emailAdd{sermakov@student.ethz.ch}
\emailAdd{klaus.kirch@psi.ch}
\emailAdd{ziehln@phys.ethz.ch}

\abstract{
The n2EDM experiment at the Paul Scherrer Institute aims to conduct a high-sensitivity search for the electric dipole moment of the neutron. Magnetic stability and control are achieved through a combination of passive shielding, provided by a magnetically shielded room (MSR), and a surrounding active field compensation system by an Active Magnetic Shield (AMS). 
The AMS is a feedback-controlled system of eight coils spanned on an irregular grid, designed to provide magnetic stability to the enclosed volume by actively suppressing external magnetic disturbances. It can compensate static and variable magnetic fields up to \SI{\pm 50}{\micro \tesla} (homogeneous components) and \SI{\pm 5}{\micro\tesla / \meter} (first-order gradients), suppressing them to a few \SI{}{\micro \tesla} in the sub-Hertz frequency range. 
We present a full finite element simulation of magnetic fields generated by the AMS in the presence of the MSR. This simulation is of sufficient accuracy to approach our measurements. We demonstrate how the simulation can be used with an example, obtaining an optimal number and placement of feedback sensors using genetic algorithms.
}


\date{}

\begin{document}
\maketitle
\flushbottom
\section{Introduction}
\label{sec:intro}
Highly stable and uniform magnetic fields are required by many fundamental physics experiments at the precision frontier to minimize systematic uncertainties, including searches for electric dipole moments, neutron oscillations, and dark matter interactions~\cite{pignol2019, Alarcon2022, Safronova2018, Jungmann2013, Mirrors2021, hibeam}. Magnetic shielding technologies benefit high-precision instruments such as electron microscopes, ion-beams, and have bio-medical applications~\cite{em_shielding, 6332797,BRYS2005527,Brake1991,Spemann2003, holmes2022}. Passive magnetic shielding is typically achieved by enclosing the experiment within layers of materials with a high magnetic permeability. Active systems often employ coils designed to generate fields capable of compensating the local residual field. If the residual field is continuously monitored by magnetic field sensors, the active system can dynamically compensate disturbance fields as they arise~\cite{shielding_overview}.

The n2EDM experiment~\cite{n2EDM}, which is currently being commissioned at the ultracold neutron (UCN) source~\cite{Lauss2012,Bison2022,Lauss2022} at the Paul Scherrer Institute (PSI), utilizes two vertically stacked UCN storage chambers. The measurement principle of the experiment relies on Ramsey's method of separated oscillating magnetic fields: a magnetic resonance technique whose sensitivity crucially relies on stable magnetic field conditions and imposes stringent limit on the vertical field gradient across the UCN storage chambers. For the same reason it is imperative that the field remains stable throughout the duration of a measurement (approximately $\SI{180}{s}$). A comprehensive discussion of the magnetic field requirements for the n2EDM experiment can be found in Ref. $\cite{n2EDM}$. The objective of the magnetic shielding design is to reduce the temporal magnetic field fluctuations inside the UCN storage chambers below $\SI{10}{\pico\tesla}$ and reduce the gradient across both UCN storage chambers to be within $\SI{0.6}{\pico\tesla\per\centi\meter}$. Both limitations are to be satisfied over the entire duration of one measurement cycle.

The n2EDM experiment meets these requirements by employing a combination of active and passive magnetic shielding. The passive shielding is provided by a magnetically shielded room (MSR)~\cite{MSR2022} with outside dimensions of $\SI{5.2}{\meter}\times\SI{5.2}{\meter}\times\SI{4.8}{\meter}$. It is constructed from seven shielding layers; six layers of mu-metal, with a maximum relative permeability of $\mu_{\rm eff} = 5\cdot10^5$, one aluminum layer for RF-shielding. The MSR provides excellent shielding, increasing from a shielding factor of $10^5$ at $\SI{0.01}{\hertz}$ up to $10^8$ at $\SI{1}{\hertz}$. The Active Magnetic Shield (AMS~\cite{AMS2023}) further stabilizes the field in the low-frequency regime. Particularly important is the stabilization of the magnetic field on the outermost layer of the MSR. This avoids changing its magnetization and further propagation of such changes through the subsequent layers to the interior volume. To the n2EDM experiment, the most important sources of magnetic disturbances are the frequent magnet ramps of the superconducting high magnetic field facility SULTAN~\cite{1060978}, and the sporadic ramps of COMET, the superconducting cyclotron of the proton therapy facility PROSCAN~\cite{SCHIPPERS2007773}.

The AMS consists of eight feedback-controlled magnet coils arranged on an irregular grid surrounding the MSR. The coils are desigend to generate the homogeneous and first-order gradient fields over the relevant volume. The AMS ensures field stability just outside of the MSR within a few $\SI{}{\micro\tesla}$ by dynamically compensating the influence of external magnetic field sources using a feedback loop based on fluxgate magnetometers. 
\begin{figure}[H]
    \centering
    \begin{subfigure}{0.48\linewidth}
        \centering
        \includegraphics[width=\linewidth]{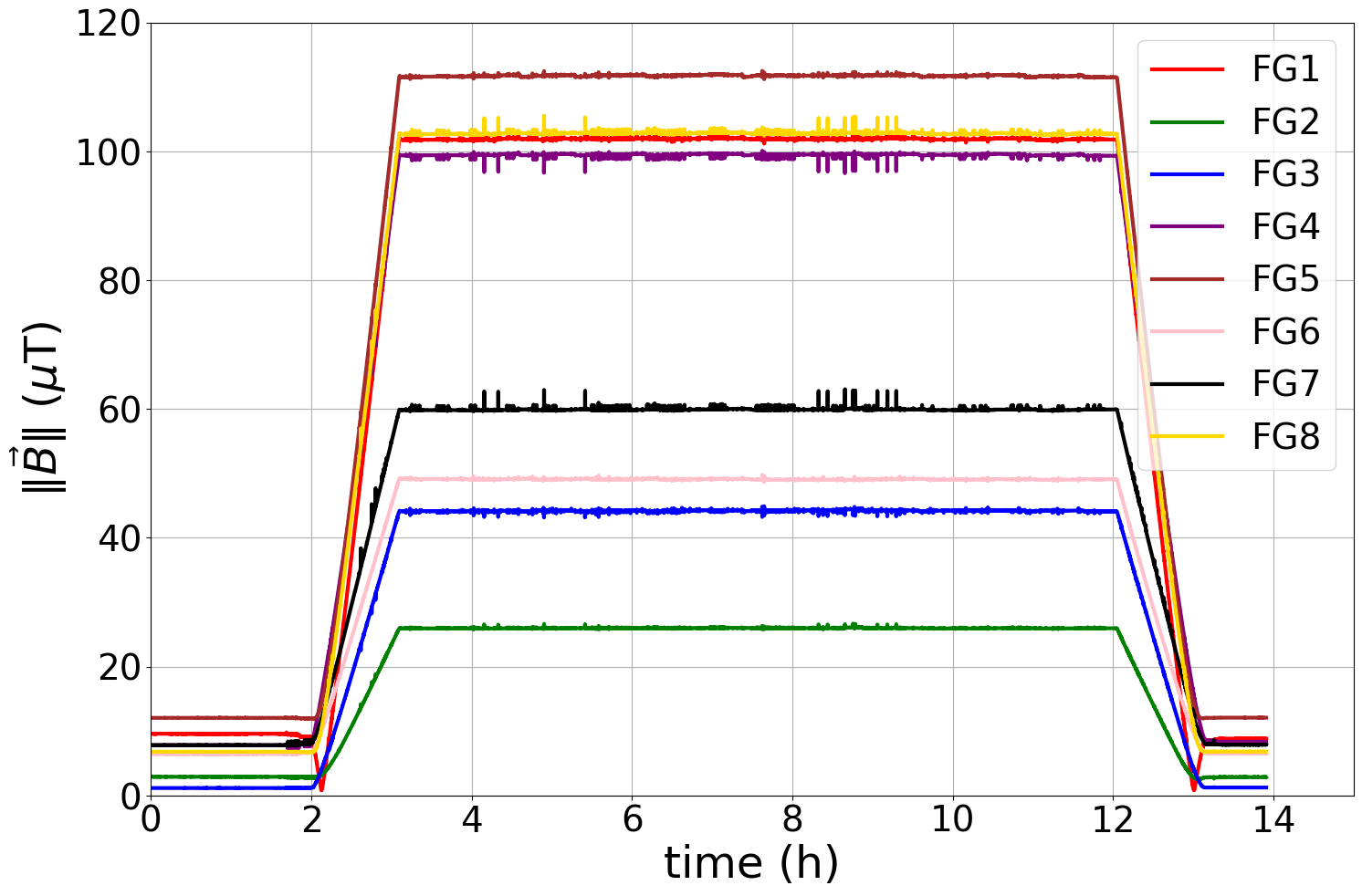}
        \caption{AMS off.}
        \label{ams_old:top}
    \end{subfigure}\hfill
    \begin{subfigure}{0.48\linewidth}
        \centering
        \includegraphics[width=\linewidth]{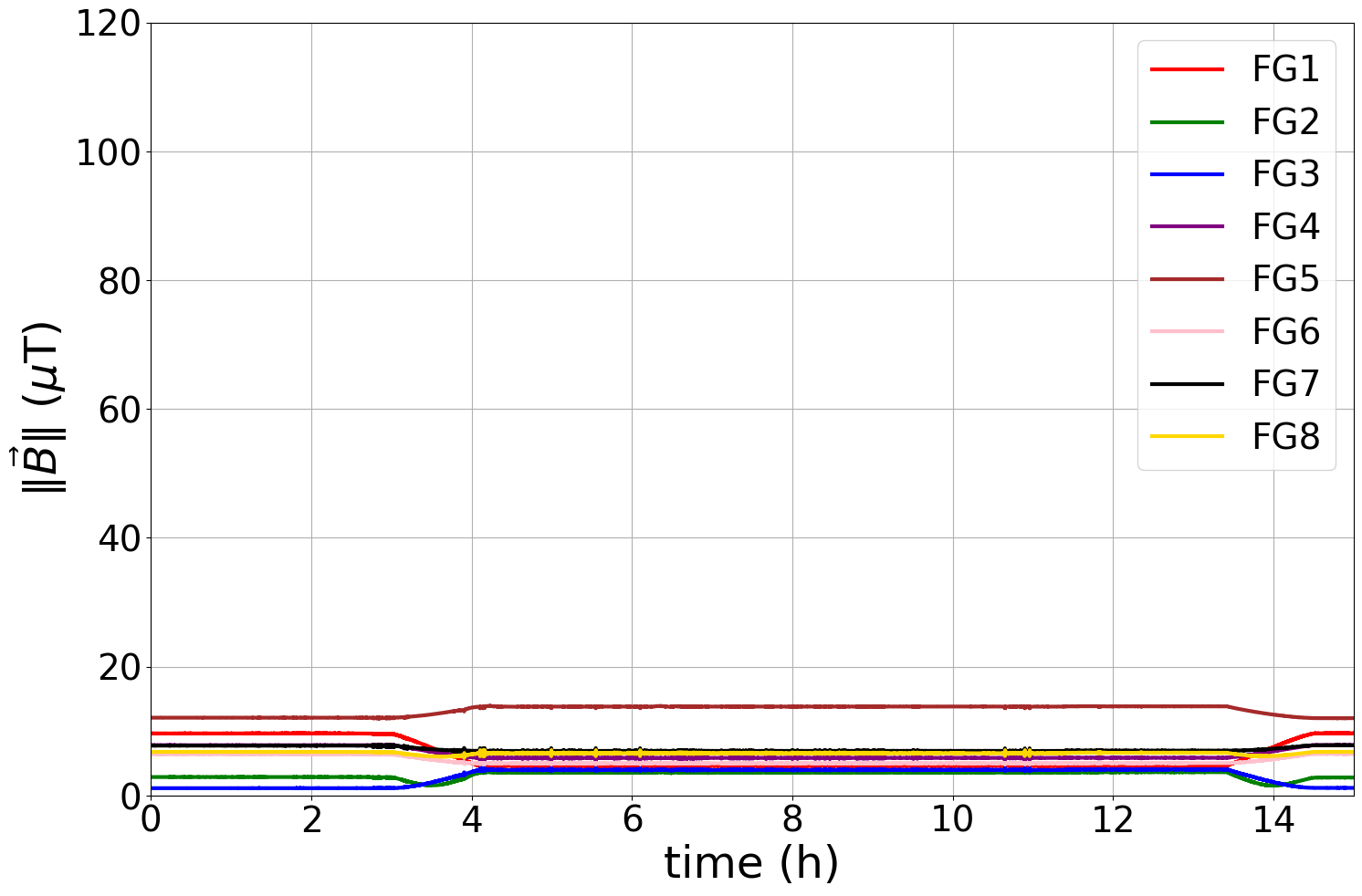}
        \caption{AMS on, for a magnified view of this plot, see Fig.~\ref{ams_performance:top}.}
        \label{ams_old:bot}
    \end{subfigure}
    \caption{Magnetic field suppression performance of the AMS after commissioning. FG1 to FG8 are used as acronyms for the respective fluxgates. The lines show the magnitude of the field picked up by fluxgate sensors when the AMS is off (\ref{ams_old:top}) and when the AMS is actively compensating the external field (\ref{ams_old:bot}), demonstrating the ability of the AMS to compensate large external field disturbances down to the level of a few $\SI{}{\micro\tesla}$. In both cases, the magnetic field change is caused by a full SULTAN ramp. Small (sub-$\SI{}{\micro\tesla}$) spikes are caused by short-term, local disturbances the AMS is not primarily designed to compensate. The placement of the sensors is discussed in section~\ref{sec:Optimization}.}
    \label{fig:ams_old}
\end{figure}

\begin{figure}[H]
    \centering
    \includegraphics[width=0.8\linewidth]{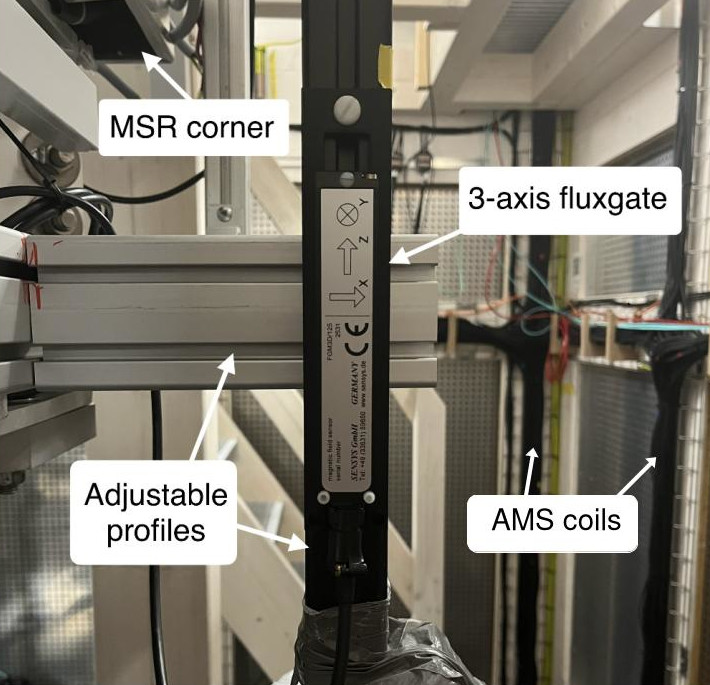}
    \caption{Image of fluxgate FG8 mounted close to a corner of the MSR.}
    \label{fig:fluxgate}
\end{figure}

The AMS was constructed and commissioned in 2021 and preliminary performance tests demonstrated its ability to suppress common magnetic field disturbances down to the level of a few $\SI{}{\micro\tesla}$~\cite{AMS2023}, as shown in Fig.~\ref{fig:ams_old}. Since then a full simulation of the system was implemented, granting a thorough understanding of the magnetic field produced by the interaction between the AMS and the MSR. In this paper we will explain how the simulation was implemented and demonstrate how it can be used to improve the performance of the AMS, using as an example the optimization of magnetic sensor placement within the AMS. To control the magnetic field, the AMS utilizes eight 3-axis  Sensys FGM3D/125 fluxgates with a range of $\pm \SI{125}{\micro\tesla}$. The sensors are mounted on non-magnetic aluminum and plastic-wood composite profiles (see e.g. Fig.~\ref{fig:fluxgate}), allowing easy adjustment of their positions by a few $\SI{}{\centi\meter}$ in all directions. The method used to obtain the initial sensor positions for commissioning the AMS was based on a gradient-descent algorithm described in Ref.~\cite{Solange2021}.

\section{AMS Design and working principle}
\subsection{Coil design}
\label{sec:coil_design}
\begin{figure}
    \centering
    \includegraphics[width=0.8\linewidth]{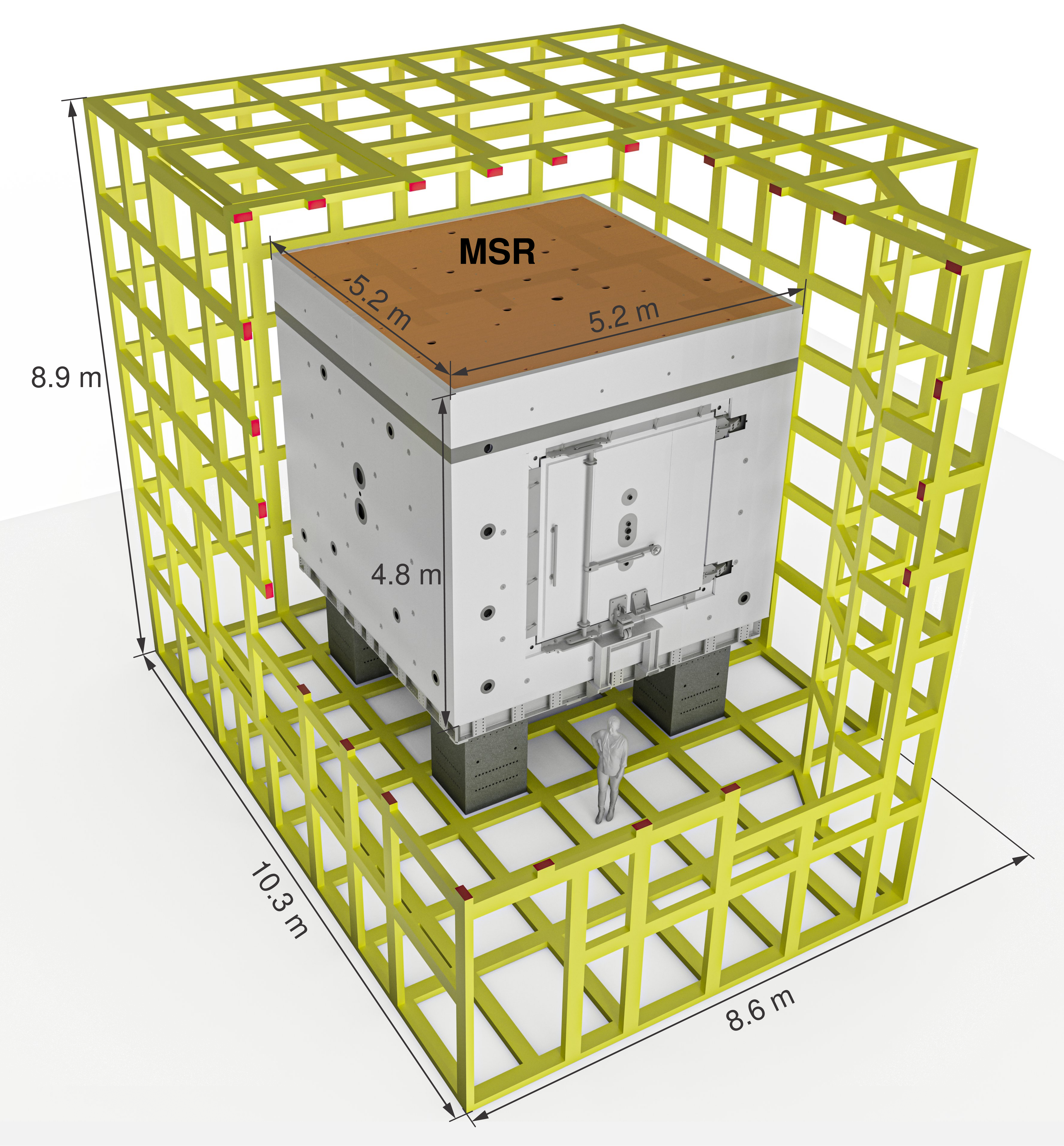}
    \caption{AMS grid structure (yellow). Parts of the grid have been cut out to allow a clear view onto the MSR. The directions of the coordinate system are indicated in the figure. The origin is located at the center of the MSR (figure adapted from~\cite{n2EDM}).}
    \label{fig:AMS_grid}
\end{figure}

Inside a volume without magnetizable parts, any source-free magnetic field can be generated by an appropriate current distribution on the surface which encloses the volume. We developed a simple method of coil design to efficiently use the limited space surrounding the MSR to generate the desired fields with suitable accuracy within the relevant volume around the Magnetically Shielded Room~\cite{Rawlik2018,Rawlik2018PhD}. This design method generates coil paths on an arbitrary, predefined grid. In the case of the n2EDM experiment, the grid was defined on the interior surface of the thermally insulating hut that encloses the apparatus. The density of the grid mesh was optimized to achieve the best possible field homogeneity on the surface of the MSR while keeping the construction efforts reasonable. The construction process is detailed in Refs. ~\cite{Solange2021, AMS2023}. The grid structure of the AMS can be seen in Fig.~\ref{fig:AMS_grid}. It measures, with some irregularities, about $\SI{10.3}{\meter} \times \SI{8.6}{\meter} \times \SI{9.8}{\meter}$ around the MSR. The rectangular tiles have an average side-length  of $\SI{1.5}{\meter}$, similar to the minimal distance of the MSR to the AMS. 

The AMS coils were designed to compensate background fields of varying shapes and sizes. Ramping of various strong magnets contributes to changing background fields in the n2EDM experimental area with homogeneous background fields of tens of $\SI{}{\micro\tesla}$ and gradients of a few $\SI{}{\micro\tesla\per\meter}$. After mapping those fields and decomposing them into basis functions, we found that it would be sufficient for the AMS to compensate up to first order gradients and we designed coils corresponding to each of these components~\cite{Solange2021}. 

Following the method described in~\cite{Franke2013}, any arbitrary source-free magnetic field may be decomposed into a sum of cartesian harmonic polynomials:

\begin{equation}
    \textbf{B}(\textbf{r}) = \sum_{n = 1}^\infty H_n \textbf{P}_n(\textbf{r})
\end{equation}
$H_n$ are the expansion coefficients and $\textbf{P}_n = (P_{nx}, P_{ny}, P_{nz})$ are the polynomials, with $n$ denoting the order. These polynomials are orthogonal and each basis state satisfies Maxwell's equations. Each AMS coil was designed to produce field shapes described by one unique polynomial: three coils generating homogeneous fields (called X-, Y- and Z-coils according to their respective field direction, see coordinate system in Fig.~\ref{fig:AMS_grid}) and five coils to compensate linear magnetic field gradients. Each homogeneous-field coil was designed to produce field strengths of up to $\pm \SI{50}{\micro\tesla}$ on the surface of the MSR, and the first order gradient coils can produce fields up to $\pm \SI{20}{\micro\tesla\per\meter}$. This configuration was sufficient to ensure that the AMS could compensate disturbance fields generated by superconducting magnets of local external facilities down to $\SI{1}{\micro\tesla}$ at the planned location. 

\subsection{Feedback mechanism}
\label{sec:feedback}

The stability of the magnetic field surrounding the MSR is ensured by a feedback mechanism. Fluxgates placed close to the exterior of the MSR continuously measure the magnetic field. Leveraging these measurements, an integral control function is used to calculate suitable adjustments to each coil current to counteract deviations from the target field.

We assumed our system would respond linearly to small magnetic field changes. We would later confirm this through measurements and simulations. The column vector $\mathbf{B}$ of all magnetic field measurements can be written as a linear combination of the currents $\mathbf{I}$ in the AMS coils plus a contribution of the background field to all measurements $\mathbf{B}_0$:
\begin{equation}
    \mathbf{B} = \mathbf{B}_0 + \mathbf{MI}.
    \label{eq:matrix_eq}
\end{equation}
$\mathbf{M}$ is the $n \times m$ dimensional matrix of measured proportionality constants that maps the coil currents to the correspondingly generated magnetic field, where $n$ is the number of individual magnetic field sensors and $m = 8$ is the number of coils. The background field $\mathbf{B}_0$ is the sum of all magnetic disturbances, such as Earth's magnetic field or from interference from other experiments. By solving the inverse problem to Eq.~\ref{eq:matrix_eq} and setting $\mathbf{B}$ equal to our target field we obtain the currents:
\begin{equation}
    \mathbf{I} = (\mathbf{B} -\mathbf{B}_0) \mathbf{M}^\dagger.
    \label{eq:inv_matrix_eq}
\end{equation}
The feedback matrix $\mathbf{M}^\dagger$ is the Moore-Penrose pseudoinverse of $\mathbf{M}$, which we calculate using singular value decomposition, as $n \neq m$:
\begin{equation}
    \mathbf{M} = \mathbf{US}\mathbf{V}^\text{T}
\end{equation}
$\mathbf{U}$ and $\textbf{V}^T$ are unitary and the spectrum $\mathbf{S}$ is a diagonal matrix whose entries $s_i$ are the singular values of $\mathbf{M}$. The pseudoinverse is:
\begin{equation}
    \mathbf{M}^\dagger = \mathbf{VS}^{-1}\mathbf{U}^\text{T}.
\end{equation}
The ratio of maximum and minimum absolute singular values yields the condition number $\sigma$ of $\textbf{M}$,

\begin{equation}\label{eq:cond}
    \sigma = \frac{|s_{\text{max}}|}{|s_{\text{min}}|}.
\end{equation}

Eq.~\ref{eq:inv_matrix_eq} is a system of $n$ linear equations to obtain $m$ parameters. For $n > m$ the problem is over-determined and the solution is found by minimizing the difference between the measured field and the target using a least-squares method. The problem is said to be well-defined if the least-squares potential around the local minimum is steep along every direction of the $m$-dimensional parameter space. 

How each current input $I_i$ contributes to the output $\mathbf{B}$ can be inferred from the matrix of singular values $\mathbf{S}^{-1}$ of $\mathbf{M}^{\dagger}$. If the spectrum is flat, all singular values are equal and $\sigma=1$. This is the case in which all input parameters contribute equally to the output of the system. Conversely, high condition numbers indicate that one or more input parameters have significantly less impact on the output than others, as described in the example above. By design the different coils of the AMS do not contribute equally to the generated magnetic field measured by the fluxgate sensors. We therefore do not expect the system to reach a condition number of $\sigma=1$. 

The surrounding field compensation employed by the n2EDM experiment's predecessor, the nEDM experiment, tackled this problem via regularization of the feedback matrix \cite{SFC2014, Franke2013}. In this approach, the singular values of $\mathbf{M}^{\dagger}$ were weighted by a regularization parameter $r$, adjusting each sensor's contribution to the applied currents. The difficulty in this approach lies in the careful tuning of $r$. 

Instead of regularizing $\mathbf{M}^{\dagger}$, the AMS obtains a good feedback matrix by modifying the sensor placements. Because $\mathbf{M}^{\dagger}$ relates the currents of the AMS to its generated magnetic field as registered by the fluxgates, the entries of $\mathbf{M}^{\dagger}$ are a direct function of sensor positions and orientations. By choosing sensor placements appropriately, the condition number of the feedback matrix $\mathbf{M}^{\dagger}$ can be minimized, thereby ensuring the stability of the feedback system without the need for regularization.

\section{COMSOL Simulation}
\label{sec:Simulation}

\subsection{Implementation}
\label{sec:Implementation}
\begin{figure}[H]
    \centering
    \includegraphics[width=0.6\linewidth]{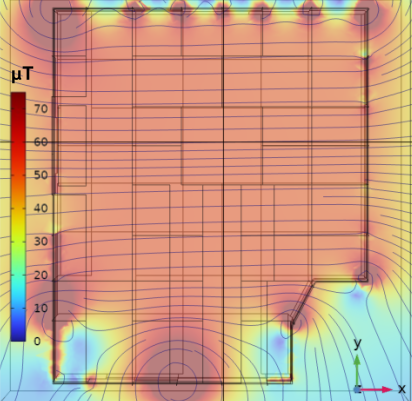}
    \caption{Simulation of the X-coil's magnetic field at maximum current, showing a top-view of an AMS cross-section at $z=0$ (plane through the middle of MSR). The simulation is performed without the MSR. The X-coil is designed to produce a homogeneous field in $x$-direction of up to $\pm \SI{50}{\micro\tesla}$. The color map indicates the magnetic field strength, gray lines represent the magnetic field lines. The wires of the X-coil are projected onto the plane of the cross-section.}
    \label{fig:COMSOL_NoMSR}
\end{figure}

The system, consisting of AMS and MSR, was modelled in COMSOL~6.1 \cite{comsol} using the Python-MPh package \cite{python_mph}. This package automates the routing of the complex wiring and current flow structures based on vertex positions and current directions of the actual AMS. For simulations, the stationary magnetic field package is used. The volume of interest, i.e. the space immediately outside of the MSR was meshed finely by tetrahedrals with sidelengths on the mm to cm scale. Larger elements were used elsewhere. A stationary magnetic field solver was employed. In order to verify that the simulation can reproduce the AMS field independent of the MSR, the AMS was simulated without the MSR. For example, the simulation of the X-coil yielded the expected homogeneous $\SI{50}{\micro\tesla}$ magnetic field at maximum current, see Fig.~\ref{fig:COMSOL_NoMSR}.

The existing MSR consists of six mu-metal layers, each with a thickness of a few millimeters. In the simulation all layers were replaced by a single $d = \SI{5}{\centi\meter}$ thick mu-metal wall to simplify the meshing. This is justified since we only consider the magnetic field outside the MSR here, for which one shell of adequate effective permeability behaves as the multilayer MSR. This is achieved by selecting $\mu_r = 2000$ cm$^{-1}$, resulting in an effective permeability of $\mu_{\text{eff}} = \mu_r\cdot d = 10000$. 

\begin{figure}[H]
    \centering
    \includegraphics[width=0.85\linewidth]{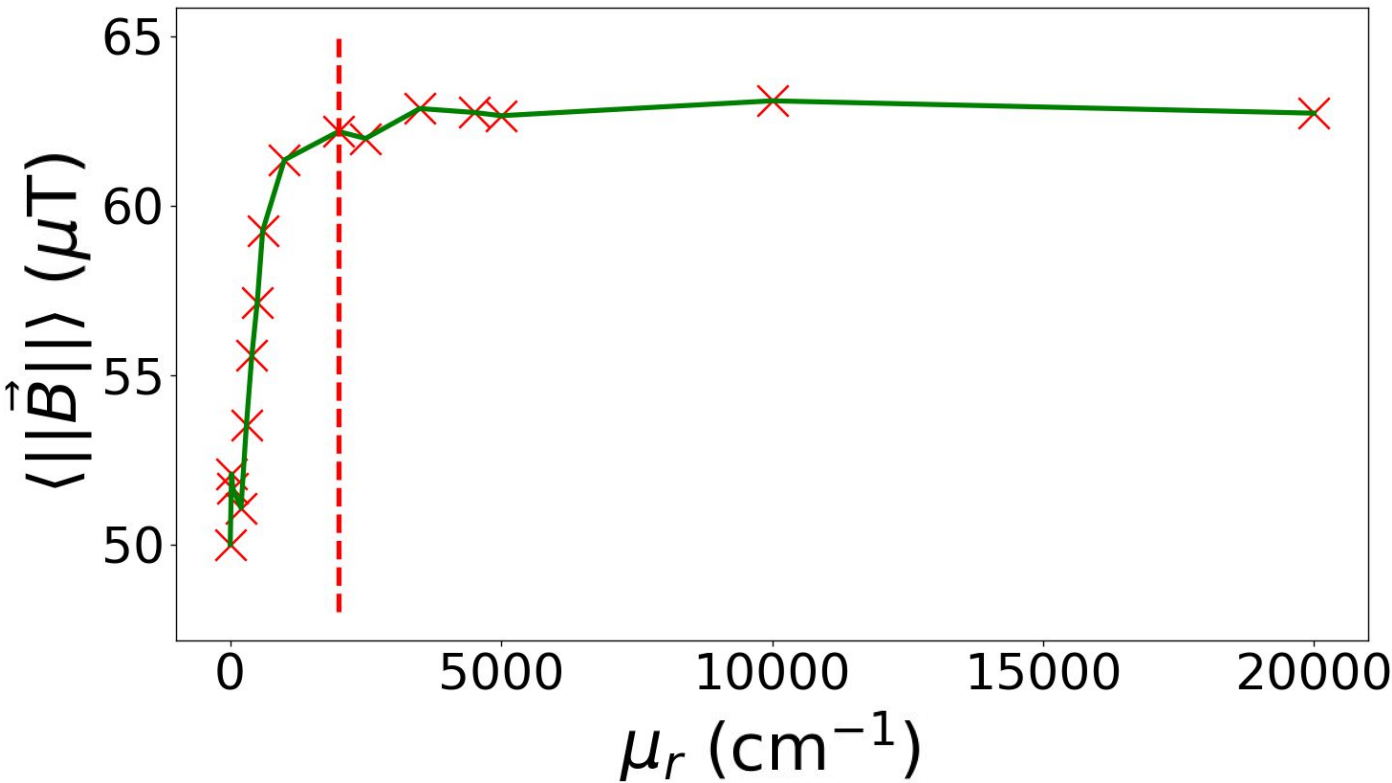}
    \includegraphics[width=0.89\linewidth]{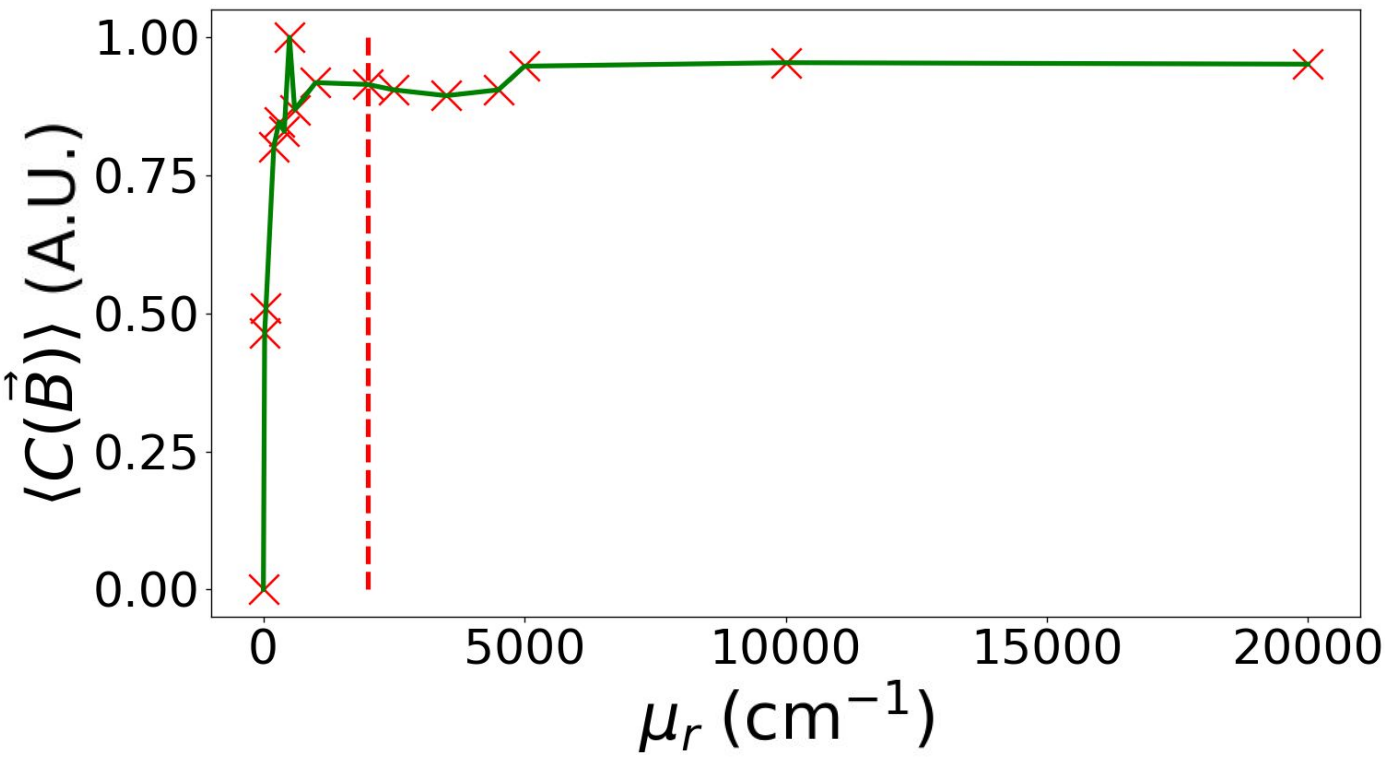}
    \caption{$\langle\|\vec{B}\|\rangle$ and $\langle C(\vec{B})\rangle$ of magnetic field lines across the AMS volume as a function of permeability of the simulated MSR. In this simulation, a $\SI{50}{\micro\tesla}$ magnetic field was applied in the $x$-direction, compare Figs.~\ref{fig:COMSOL_NoMSR} and~\ref{fig:COMSOL_MSR}. A vertical red line indicates where $\mu_r=2000~\rm cm^{-1}$.}
    \label{fig:VaryingMu}
\end{figure}

In order to study the effect of $\mu_{r}$ on the surrounding magnetic field, we embedded the MSR in a $\SI{50}{\micro\tesla}$ homogeneous, external magnetic field in the $x$-direction for various values of $\mu_{\text{eff}}$, as shown in Fig.~\ref{fig:VaryingMu}. For this, we create a new simulation, with approximately constant mesh size across the full AMS volume. To assess the influence of $\mu_{\text{eff}}$ on the magnetic field, we evaluate two quantities: the mean absolute magnetic field strength and the mean curvature of the magnetic field lines. The finite element simulation discretizes space into mesh points, so the spatial average of the field magnitude is given by
\begin{equation}
    \langle\|\vec{B}\|\rangle = \frac{1}{N_{\text{mesh}}}\sum_{k\in\text{mesh}}\|\vec{B}(\vec{r}_k)\|,
    \label{eq:b_av}
\end{equation}
where $N_{\text{mesh}}$ is the total number of mesh points outside the MSR.  
Magnetic field lines are discretized as parametrized curves $\vec{C}(t)$ defined on these mesh points. The average curvature of all field lines is then given by~\cite{Cartan1983Geometry}
\begin{equation}
    \langle C(\vec{B})\rangle = \frac{1}{N_{\text{curve}}}\sum_{k\in\text{curves}} \frac{\|\dot{\vec{C}}(t_k)\times\ddot{\vec{C}}(t_k)\|}{\|\dot{\vec{C}}(t_k)\|^3}.
    \label{eq:c_av}
\end{equation}
Derivatives $\dot{\vec{C}}$ and $\ddot{\vec{C}}$ are obtained numerically from the discrete field-line points. The average magnetic field strength and the curvature were studied as a function of $\mu_r$ by performing simulations for some chosen values of $\mu_r$. 
Starting from $\mu_r = 2000$, the values for $\langle\|\vec{B}\|\rangle$ and $ \langle C(\vec{B})\rangle$ plateau. Above this threshold, deviations in $\mu_{\text{eff}}$ have no significant effect on the outside magnetic field. Therefore the effective permeability of our simulation does not need to match the $\mu_{\text{eff}}$ of the actual MSR exactly\footnote{We note in passing that the measurement of the shielding factor of the single outermost layer of the MSR during construction resulted in an effective $\mu_{\rm eff} \approx 24'000$ for this layer, corroborating our choice.}, as long as both values are sufficiently large.
\begin{figure}[H]
    \centering
    \includegraphics[width=0.75\linewidth]{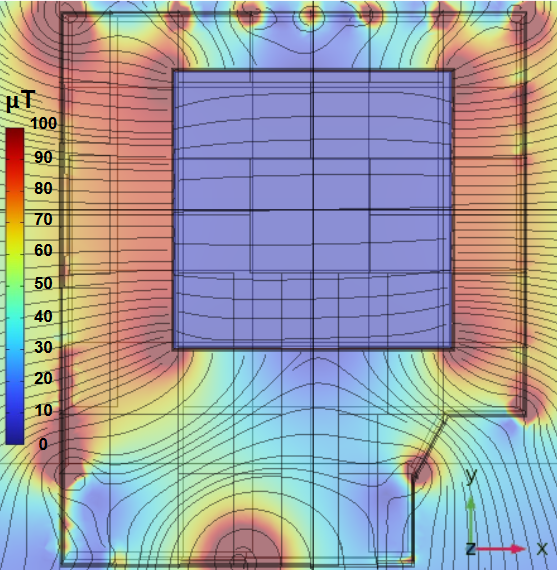}
    \caption{Same cross-section view as Fig.~\ref{fig:COMSOL_NoMSR}. Simulation of the AMS and the MSR showing the magnetic field generated by the X-coil at maximum current and distorted by the MSR. The conservation of magnetic flux results in an amplification of the magnetic field across the $yz$-surface surrounding the MSR, compared to Fig.~\ref{fig:COMSOL_NoMSR}. The colormap shows the magnetic field magnitude with corresponding field lines.}
    \label{fig:COMSOL_MSR}
\end{figure}
The high permeability of the mu-metal distorts the magnetic field of the homogeneous X-coil, as quantified in Fig.~\ref{fig:VaryingMu}, with particularly high magnetic field values observed at the corners and edges of the MSR (see Fig.~\ref{fig:COMSOL_MSR}).

\subsection{Experimental Verification}
\label{sec:Experimental-verification}
\begin{figure}[H]\label{fig:CompareExperimentToSimulation}
    \centering
    \includegraphics[width=0.75\linewidth]{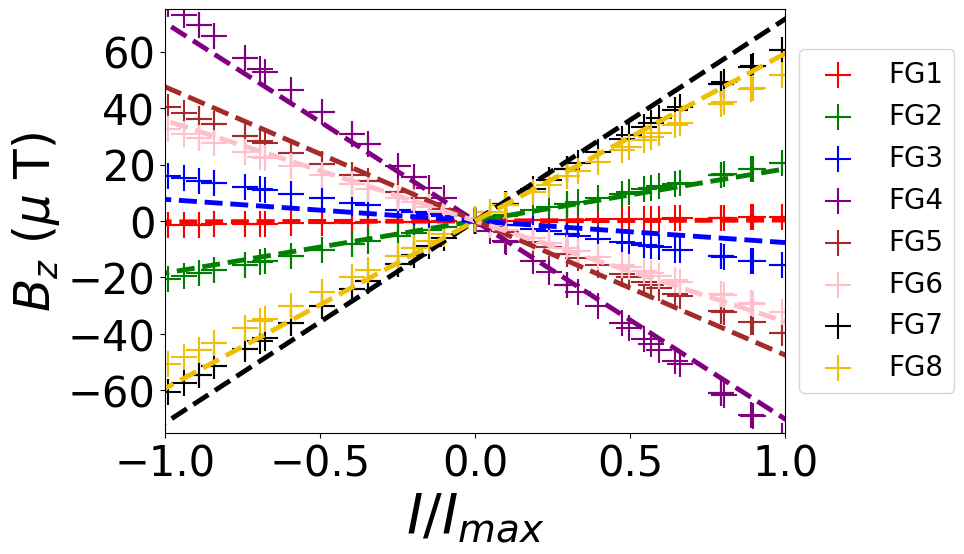}
    \caption{Comparison between simulation (dashed lines) and experimental measurements (crosses) of the z-component of the magnetic field generated by the homogeneous X-coil as a function of normalized input current. Each color represents a different sensor. The background magnetic field in the experimental area was measured at zero current and subtracted from the measurement data to eliminate any vertical offsets between the two datasets. While FG7 deviates by approximately $10~\mu \rm T$, all other fluxgates deviate by at most $5~\mu \rm T$ at maximum current, $I_{\rm max}$.}
    \label{fig:SimulationVsExperiment}
\end{figure}

We verify the simulation by comparing the coil-current dependence of the magnetic field to measurements from the live system. Both simulation and experiment show the expected linear dependence. The proportionality constants $m_{ij}$, as defined in section~\ref{sec:feedback}, were determined by ramping each AMS coil individually from its minimum to maximum current and recording the resulting magnetic-field changes with the fluxgate sensors. The same was done in the simulation, allowing for a comparison between measured and simulated data at all fluxgate positions, see Fig.~\ref{fig:SimulationVsExperiment} for a full ramp of the X-coil.

The simulation shows good agreement with the measurements, as seen in Fig.~\ref{fig:SimulationVsExperiment}. We quantify the uncertainty due to inaccuracy in measured fluxgate position up to 2.5~cm by sampling a random displacement within $\pm 2.5~\rm cm$ around the nominal fluxgate positions, shown in Fig.~\ref{fig:Configurations} on the top, and repeating the linear fit shown in Fig.~\ref{fig:SimulationVsExperiment}. We find that one standard deviation of the slope value is at most $0.61\mu \rm T/\rm I_{max}$ which is negligible.   

As discussed in section~\ref{sec:feedback}, the condition number of the feedback matrix is an important diagnostic metric. We verify agreement between the simulated condition number $\sigma_{\text{sim}} = 10.41$ and the measured condition number $\sigma_{\text{exp}} = 10.18$ at the nominal fluxgate positions. As in the comparison with measured magnetic field values, the condition number contains an uncertainty stemming from inaccuracy of fluxgate positions. To quantify this, we again sample condition numbers from our simulation, at random displacements within $\pm 2.5\rm cm$. We find a mean value of $\sigma_{\rm sim}=11.41\pm 0.9$ where the uncertainty corresponds to one standard deviation. The experimental condition number agrees within $1.37$ standard deviations. An additional systematic uncertainty arises from small, unquantified displacements of the MSR relative to its nominal position, as well as minor deviations of its dimensions from the design specifications. 

From Fig.~\ref{fig:SimulationVsExperiment} it is also evident that the relationship between the current running through the AMS coils and the measured magnetic field is linear. This verifies the initial assumption of linearity made in section~\ref{sec:feedback}.

The experimental data shows that the AMS does not magnetize the MSR, unlike suspected previously~\cite{Solange2021}. This can also be shown by including a magnetization curve in the same COMSOL simulation as in Fig.~\ref{fig:COMSOL_MSR}, which yields the same slopes as in Fig.~\ref{fig:SimulationVsExperiment}. While the magnetization curve of the mu-metal included in COMSOL might not exactly correspond to that of our system, this is irrelevant as we merely show that the magnetic field range covered by the AMS coils is small enough for magnetization effects to be negligible.

Due to the linearity of the system, further simulations of the AMS can be reduced to a linear algebra problem. The system does not have to be simulated in COMSOL for each current value, but can instead be exported at a fixed reference current and scaled with the current vector according to Eq.~\ref{eq:matrix_eq}. This allows for many applications, including a simple simulation of the compensation for external magnetic field disturbances, as well as the optimization of the number of fluxgates and their positions.

\subsection{Simulation of Compensation}
\label{sec:Explained-spread}
The AMS design goal is to suppress external magnetic fields down to $\SI{1}{\micro\tesla}$ in each direction without the MSR. Since the fluxgates designed to monitor the residual field are primarily positioned at the MSR corners, where the magnetic field is the strongest due to the distortion by the mu-metal, an amplification of the measured residual field is expected (see Fig.~\ref{fig:COMSOL_MSR}).

\begin{figure}[H]
    \centering
    \includegraphics[width=0.75\linewidth]{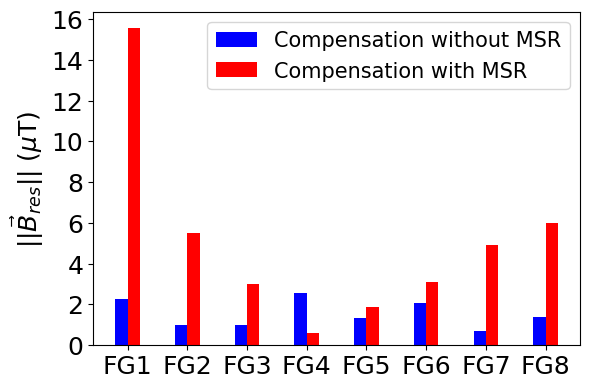}
    \caption{Residual field strength (absolute values) after AMS compensation of a homogeneous $\SI{50}{\micro\tesla}$ magnetic field in the $x$-direction, evaluated at each fluxgate position for a simulation with (red) and without the MSR (blue).}\label{fig:WithWithoutMSR}
\end{figure}

\begin{figure}[H]
    \centering
    \includegraphics[width=0.6\linewidth]{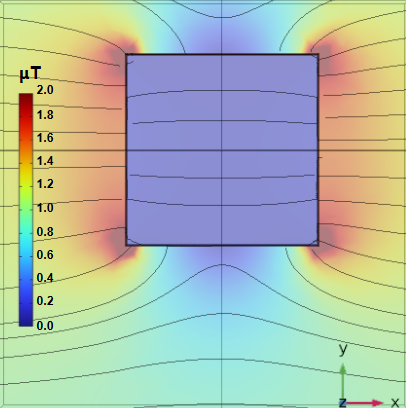}
    \caption{Simulation showing the distortion of a $\SI{1}{\micro\tesla}$ homogeneous magnetic field by the mu-metal. Due to the linearity of the system, the input may conveniently be scaled up to any external magnetic field value. Here we choose to scale the disturbance to $\SI{50}{\micro\tesla}$ magnitude. Note that the highest amplification takes place at the corners. The same cross-section view as for Figs.~\ref{fig:COMSOL_NoMSR} and~\ref{fig:COMSOL_MSR} is shown.}\label{fig:1muTCOMSOL}
\end{figure}

To quantify the amplification, we simulate the compensation of an external disturbance, with and without the MSR in place. We consider a uniform, external magnetic field as a disturbance, as shown in Fig.~\ref{fig:1muTCOMSOL}. We export the magnetic field of each coil for both simulations with and without MSR. The magnetic fields can then be scaled linearly as described for the AMS fields. Subsequently, a least-squares algorithm was applied to the superposition of the exported magnetic fields in python, calculating the compensation currents to apply and thereby simulating the AMS control algorithm. The compensation currents were calculated separately with and without the MSR. The simulation results, evaluated at each fluxgate position, are shown in Fig.~\ref{fig:WithWithoutMSR}.
This demonstrates that the capability of the simulation to reproduce the AMS compensating the disturbance field down to a few $\mu\text{T}$ at the chosen fluxgate positions in the absence of the MSR. When the MSR is introduced, the simulation reproduces the residual field strength increase above $\SI{10}{\micro\tesla}$, depending on the individual fluxgate's position. This behavior is consistent with observations made on the working system, where remnant fields up to $\SI{10}{\micro\tesla}$ remain despite suppression of the disturbances by the AMS, see Fig.~\ref{fig:ams_old}.

\section{An Example for Optimization}
\label{sec:Optimization}

\subsection{Motivation and Genetic Algorithms}
\label{sec:Genetic-algorithms}
With the simulation verified against experimental results, we may exploit its linear behavior for various applications. We present one of many possible optimization examples. Here, we optimize the fluxgate configuration based on optimizing the condition number, introduced in Eq.~\ref{eq:cond}. We optimize for the best number of sensors, their respective position and orientation in space. This presents a high-dimensional problem, since we have the number of fluxgates $n$ with $3$-dimensional position space and two angles (azimutal and polar) introducing a $6n$-dimensional parameter space.

Genetic algorithms (GAs) are popular stochastic optimization methods inspired by natural selection, commonly used for high dimensional optimization problems~\cite{goldberg1989,hollland1975}. GAs have been successfully applied to sensor placement in control systems~\cite{haupt2004}, electromagnetic shielding design~\cite{liu2008} and magnetic field optimization~\cite{Pais2018, miranda2010}. 

We note that by placing the fluxgates in the corners we can therefore expect them to be most sensitive to the AMS coils, since the magnetic field values are highest here. Additionally, we expect that suppressing the field in these sections will also result in the compensation in regions of lower magnetic field strength. As we are only interested in controlling the magnetic field close to the MSR, besides minimizing the condition number, the optimization also focuses on minimizing the average distance of fluxgate sensors to the MSR corners. Thus, our problem becomes a multi-objective optimization problem. Introducing this second parameter forces the genetic algorithm to converge to values optimal to both condition number and corner distance, forming a Pareto front of solutions. 

In multi-objective optimization a single global optimum is rare; instead, a set of solutions, each representing a trade-off, forms the Pareto front, illustrating the best possible outcomes across the two optimization targets. For a detailed explanation of the genetic algorithm see Ref.~\cite{akiba2019optuna}. 

Different numbers of fluxgates, orientations, and positions of the fluxgates are explored, with the algorithm evaluating the condition number and distance to MSR corners for each setup in every iteration. This gradually evolves them toward an optimal arrangement.

\subsection{Optimization Results}
\label{sec:Optimization-results}
\begin{figure}[H]
    \centering
    \includegraphics[width=0.7\linewidth]{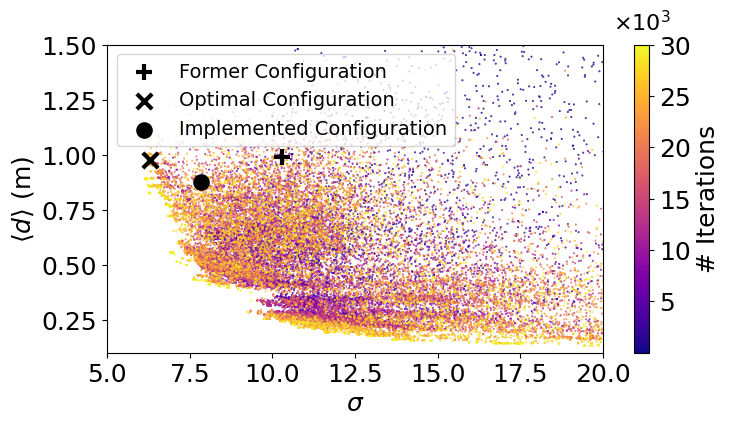}
    \caption{Condition number $\sigma$ and average distance to the nearest MSR corners $\langle d\rangle$ for each configuration across genetic algorithm iterations. The Pareto front is marked by yellow values from the highest iteration, with key points highlighted by black markers.}
    \label{fig:ParetoFront}
\end{figure}

The Pareto front of the optimization process is shown in Fig.~\ref{fig:ParetoFront}. A key observation is a characteristic point where the condition number reaches approximately 7.5, and the average distance of each sensor to their nearest MSR corners is $\SI{0.4}{\meter}$. Moving the sensors any closer to the corners leads to a significant rise in the condition number. Since the primary objective was to minimize $\sigma$, several configurations with an average MSR distance larger than $\SI{0.4}{\meter}$ were also evaluated, the best one being indicated by a cross. Most of them show $8$ fluxgates as the optimal number of fluxgates. 

\begin{figure}[H]
    \centering
    \begin{subfigure}[a]{\linewidth}
        \centering
        \includegraphics[width=0.65\linewidth]{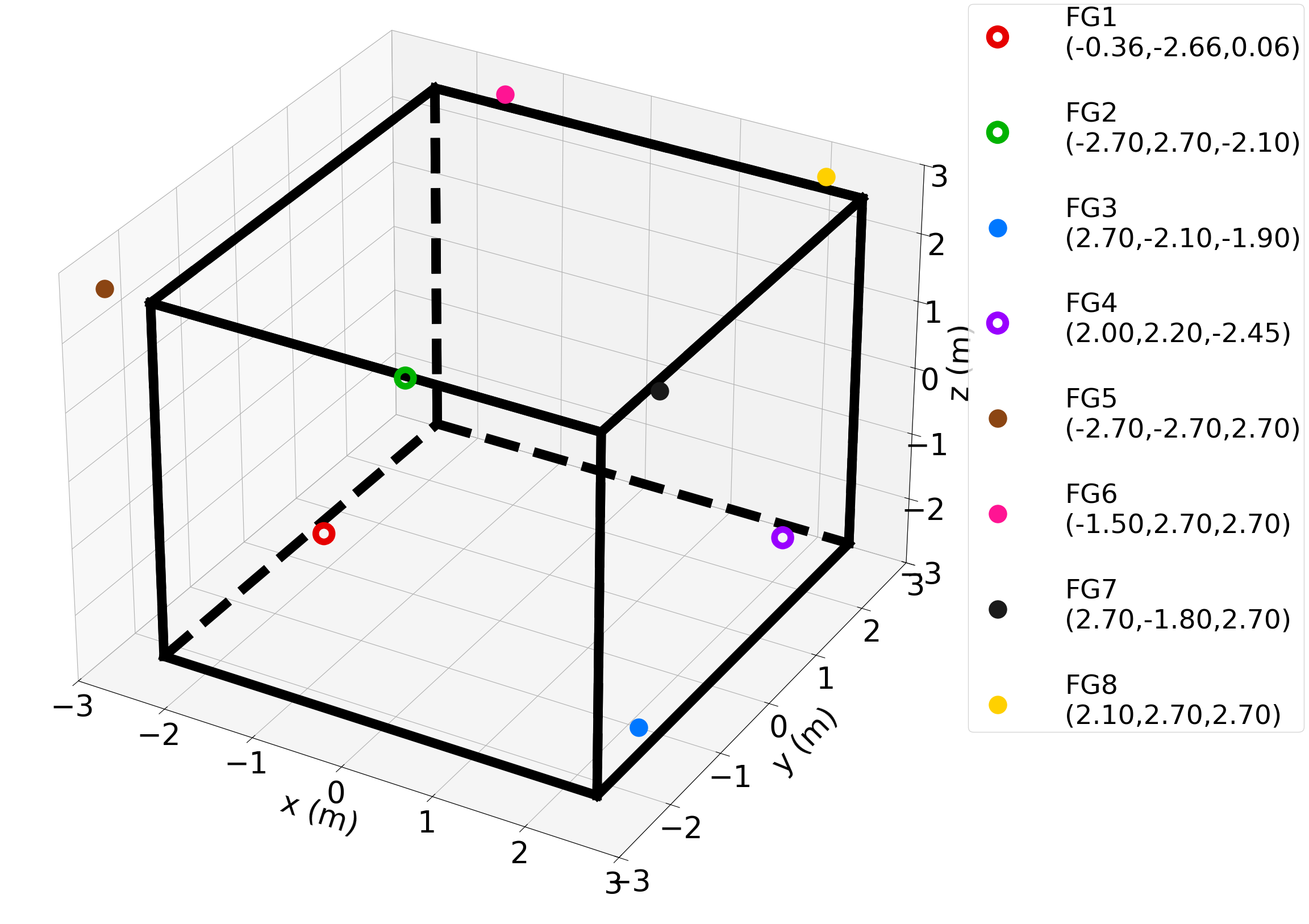}

        \caption{Former fluxgate configuration.}
        \label{ams_performance:top}
    \end{subfigure}
    \begin{subfigure}[b]{\linewidth}
        \centering
        \includegraphics[width=0.65\linewidth]{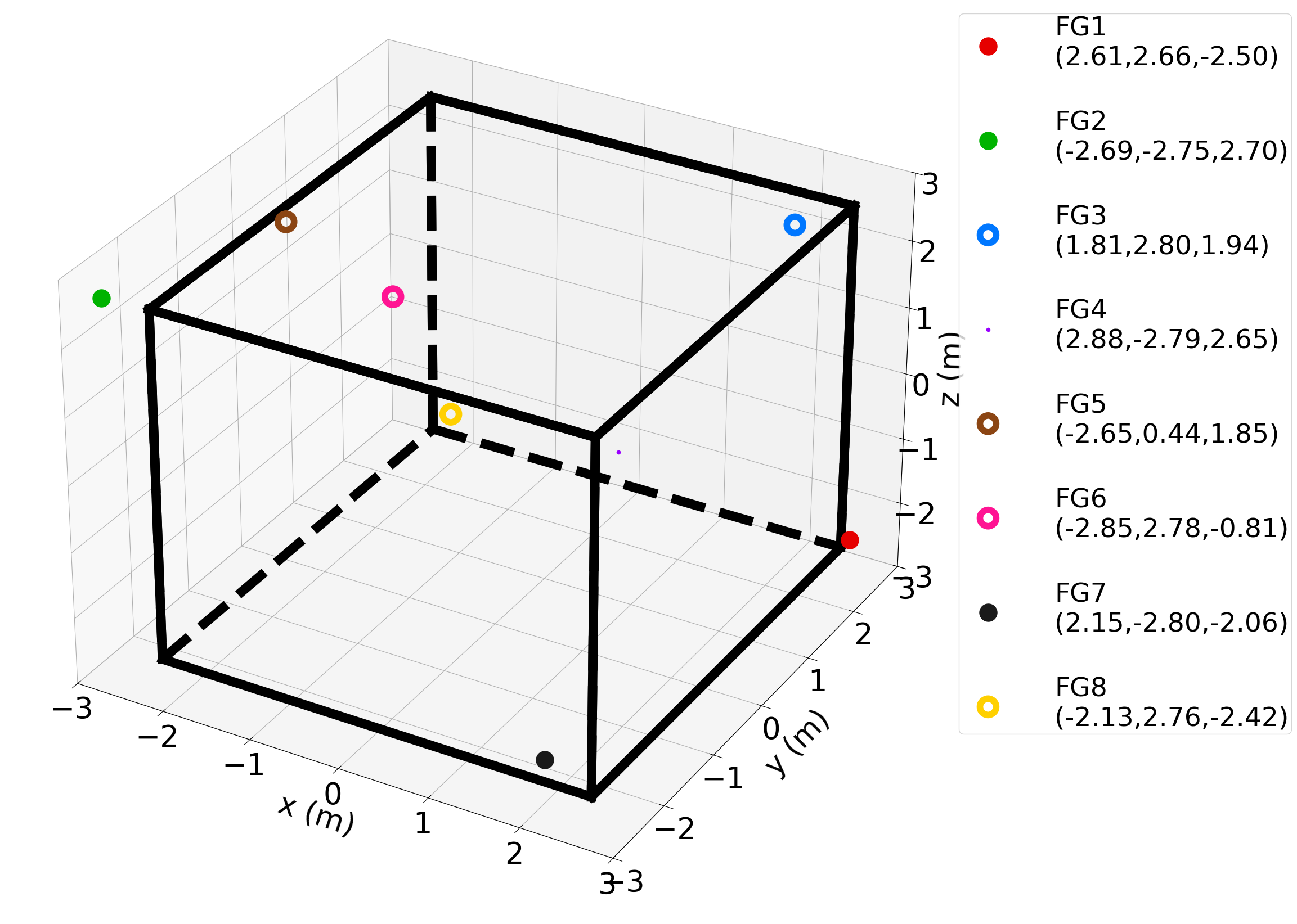}

        \caption{Implemented fluxgate configuration.}
        \label{ams_performance:bot}
    \end{subfigure}

    \caption{Comparison of fluxgate positions, compare Fig~\ref{fig:ParetoFront}: The former configuration is shown in the top figure, the newly implemented configuration is shown in the bottom figure. The coordinates are given in units of meters. Fluxgates behind or below the MSR are represented by a hollow symbol.} 
    \label{fig:Configurations}
\end{figure}

The optimal configuration was impractical due to limited space around the experiment. The best condition number achieved by the genetic algorithm was $\sigma = 6.3$, while the adjusted configuration, accounting for physical constraints, resulted in $\sigma = 7.85$. This alternative setup, marked by a black dot in Fig. \ref{fig:ParetoFront}, was ultimately selected. Both configurations are shown in Fig. $\ref{fig:Configurations}$.

\subsection{Experimental Implementation}
\label{sec:Experimental-implementation}
An experimental verification of the optimal condition number was conducted by implementing the fluxgate positions chosen in section~\ref{sec:Optimization-results} on the working system.

Given the difficulty in accurately measuring fluxgate positions, fluxgates were initially placed at positions corresponding to the expected distances from their nearest MSR corner. Then each fluxgate was moved by up to $\SI{4}{\centi\meter}$ in each direction in order to find the configuration that best matches the optimal arrangement (see section~\ref{sec:Experimental-verification}). The best configuration yielded $\sigma_{exp} = 8.15$, close to the simulated condition number of $\sigma_{sim} = 7.85$, validating the optimization workflow.

\begin{figure}[H]
    \centering
    \begin{subfigure}{0.48\linewidth}
        \centering
        \includegraphics[width=\linewidth]{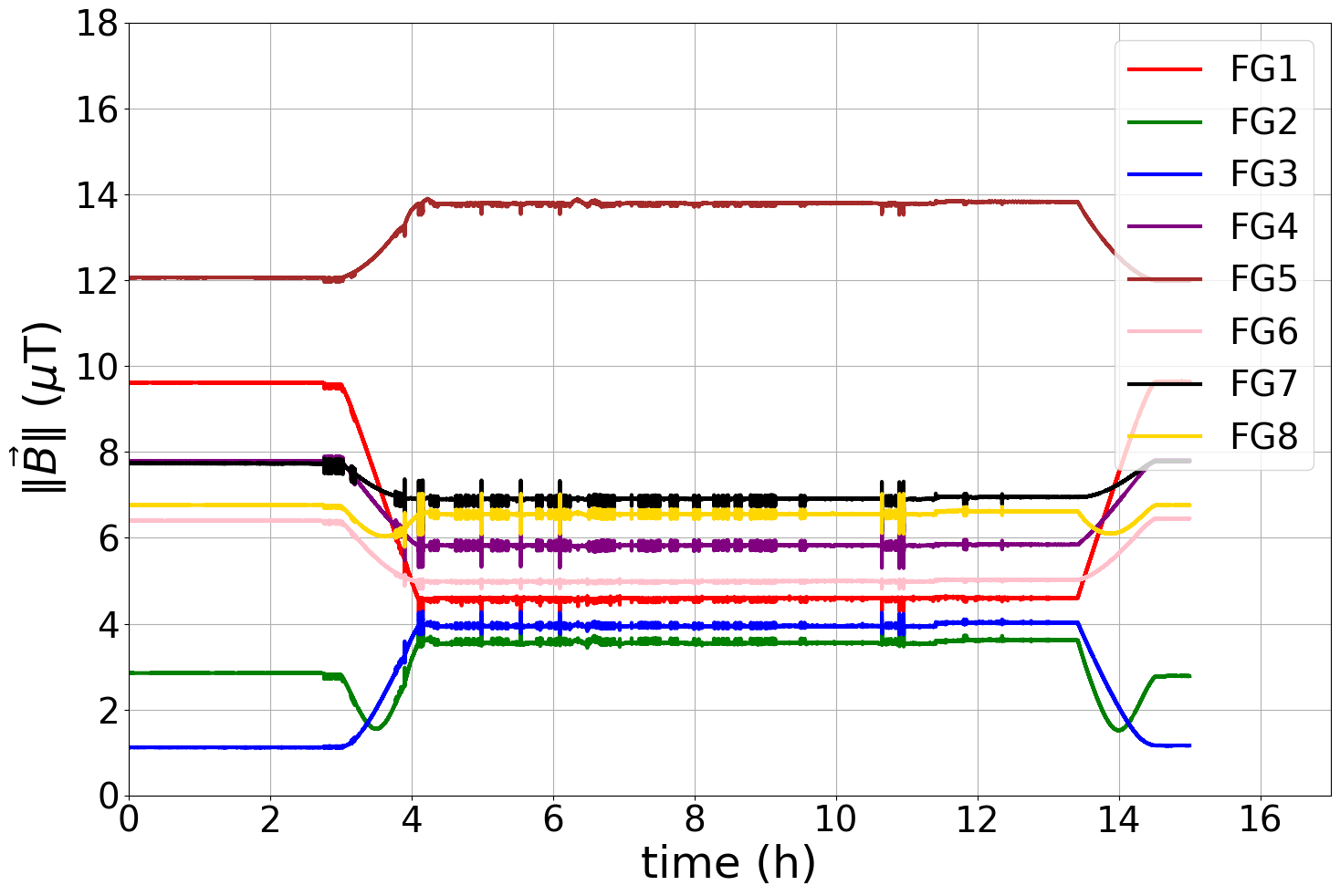}
        \caption{Old fluxgate configuration.}
        \label{ams_performance:top}
    \end{subfigure}\hfill
    \begin{subfigure}{0.48\linewidth}
        \centering
        \includegraphics[width=\linewidth]{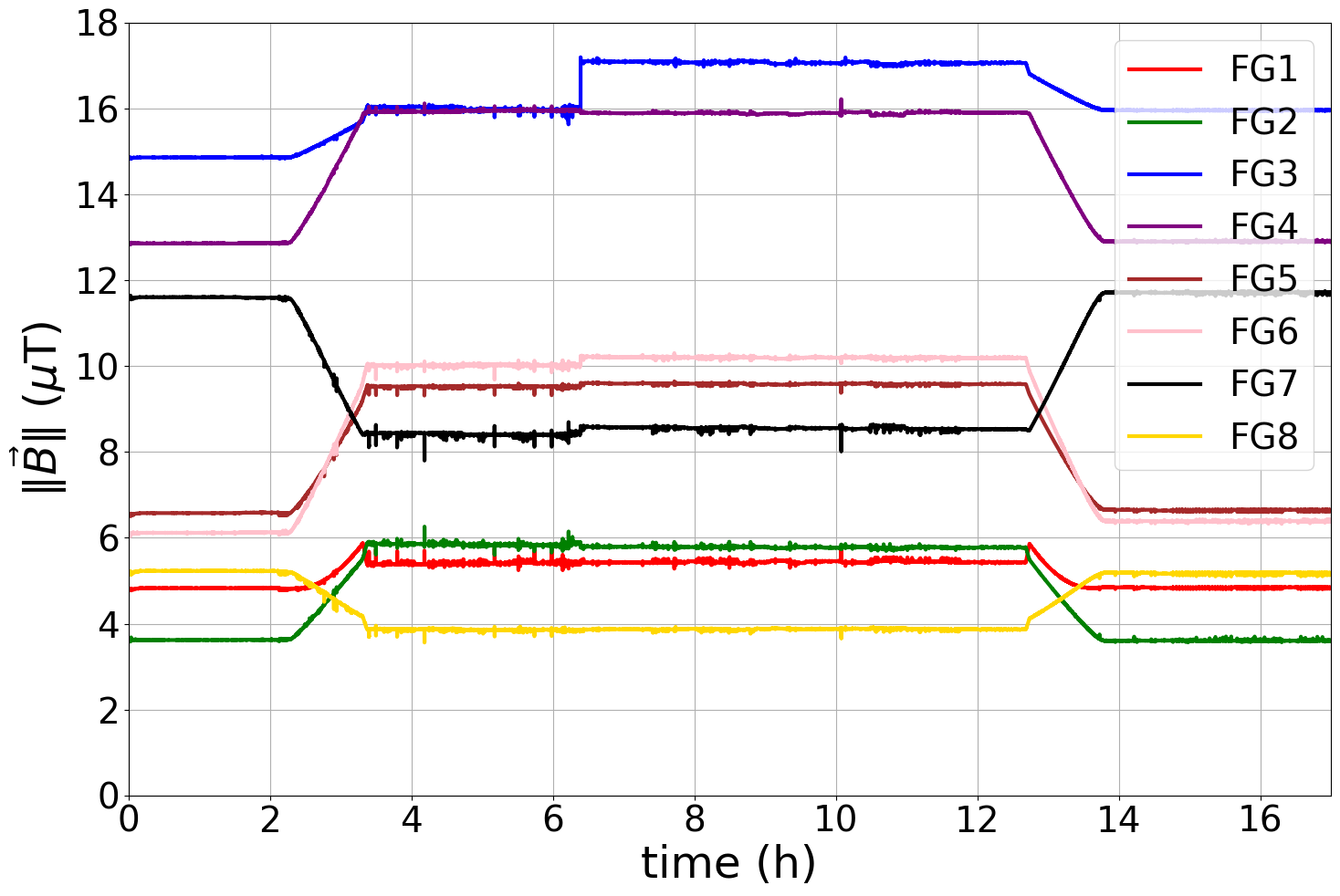}
        \caption{New fluxgate configuration.}
        \label{ams_performance:bot}
    \end{subfigure}
    \caption{Comparison of the former fluxgate configuration in operation (\ref{ams_performance:top}) and the newly implemented configuration (\ref{ams_performance:bot}), compensating a SULTAN ramp between hours 3 and 14.}
    \label{fig:ams_performance}
\end{figure}

The new configuration yielded slightly higher residual magnetic fields than the former configuration, as shown in Fig.~\ref{fig:ams_performance}. This is due to the fact that the new fluxgate positions are closer to MSR corners, where the magnetic field is stronger.

\section{Conclusion and Outlook}
\label{sec:Conclusion}
We presented a detailed simulation of the active magnetic shielding system of the n2EDM experiment that takes into account the magnetic field distortions caused by the magnetically shielded room. The simulation improved the understanding of the whole system. We were able to verify that the AMS meets its design objectives, as outlined in~\ref{sec:coil_design}, and to understand how the mu-metal changes the residual fields surrounding the MSR. While enhancing the magnitude of the magnetic field at its corners, the MSR does not affect the linear dependence between AMS coil currents and generated magnetic fields. This confirms that a linear control algorithm can pilot the AMS.

We successfully employed genetic algorithms to optimize the number and positions of feedback sensors. By requiring the sensors to be in close proximity to the MSR corners and optimizing for the condition number of the feedback matrix, we ensure that the system is fully controllable and stable. 

In future works, other parameters of the AMS can be optimized for, e.g. the residual magnetic field. Other properties of the AMS can be examined through the simulations such as self-inductance of the AMS coils. 

The developed workflow for optimizing the condition number may be applied to other hardware and to active magnetic shields in other experiments.

\acknowledgments
Excellent technical support by Michael Meier and Luke Noorda is acknowledged. Various PSI LOG groups supported the electrical components construction.

Support by the Swiss National Science Foundation Projects 188700 (PSI), 163413 (PSI), 178951 (PSI), 204118 (PSI), 172626 (PSI), 213222 (PSI), 169596 (PSI), 163663 (Bern), 181996 (Bern), 215185 (Bern), 236419 (ETH), 200441 (ETH), 10003932 (ETH), and FLARE 186179, 201473, 216603, 232701 are gratefully acknowledged.

This project has received funding from the European Union’s Horizon 2020 research and innovation programme under the Marie Skłodowska-Curie grant agreement No 884104 and No 701647

This work is support by the DFG (DE) by the funding of the PTB core facility center of ultra-low magnetic field KO 5321/3-1 and TR 408/11-1.

The LPC Caen and the LPSC Grenoble acknowledge the support of the French Agence Nationale de la Recherche (ANR) under reference ANR-14-CE33-0007 and the ERC project 716651-NEDM.

University of Bern acknowledges the support via the European Research Council under the ERC Grant Agreement No. 715031-Beam-EDM.

The Polish collaborators wish to acknowledge support from the National Science Center, Poland, under grant No. 2018/30/M/ST2/00319, and No. 2020/37/B/ST2/02349, as well as by the Minister of Education and Science under the agreement No. 2022/WK/07.

Support by the Cluster of Excellence ‘Precision Physics, Fundamental Interactions, and Structure of Matter’ (PRISMA+ EXC 2118/1) funded by the German Research Foundation (DFG) within the German Excellence Strategy (Project ID 39083149) is acknowledged.

University of Sussex collaborators wish to acknowledge support from the School of Mathematical and Physical Sciences, as well as from the STFC under grants ST/S000798/1 and ST/W000512/1.

This work was partly supported by the Fund for Scientific Research Flanders (FWO), and Project GOA/2010/10 of the KU Leuven.


\bibliographystyle{JHEP}
\bibliography{Zotero.bib}

\providecommand{\href}[2]{#2}\begingroup\raggedright\begin{thebibliography}{10}

\bibitem{pignol2019}
G.~Pignol, \emph{A global perspective on searches for {Electric} {Dipole}
  {Moments}}, \href{https://doi.org/10.48550/arXiv.1912.07876}{\emph{arXiv}
  (2019) }.

\bibitem{Alarcon2022}
R.~Alarcon, J.~Alexander, V.~Anastassopoulos, T.~Aoki, R.~Baartman, S.~Baeßler
  et~al., \emph{Electric dipole moments and the search for new physics},
  \href{https://doi.org/10.48550/arXiv.2203.08103}{\emph{arXiv} (2022) }.

\bibitem{Safronova2018}
M.S.~Safronova, D.~Budker, D.~DeMille, D.F.J.~Kimball, A.~Derevianko and
  C.W.~Clark, \emph{Search for new physics with atoms and molecules},
  \href{https://doi.org/10.1103/RevModPhys.90.025008}{\emph{Rev. Mod. Phys.}
  {\bfseries 90} (2018) 025008}.

\bibitem{Jungmann2013}
K.~Jungmann, \emph{Searching for electric dipole moments},
  \href{https://doi.org/10.1002/andp.201300071}{\emph{Ann. Phys.} {\bfseries
  525} (2013) 550}.

\bibitem{Mirrors2021}
C.~Abel, N.~Ayres, G.~Ban, G.~Bison, K.~Bodek, V.~Bondar et~al., \emph{A search
  for neutron to mirror-neutron oscillations using the {nEDM} apparatus at
  {PSI}},
  \href{https://doi.org/https://doi.org/10.1016/j.physletb.2020.135993}{\emph{Physics
  Letters B} {\bfseries 812} (2021) 135993}.

\bibitem{hibeam}
V.~Santoro, D.~Milstead, P.~Fierlinger, W.M.~Snow, J.~Barrow, M.~Bartis et~al.,
  \emph{The {HIBEAM} program: search for neutron oscillations at the {European
  Spallation Source}}, \href{https://doi.org/10.1088/1361-6471/adc8c2}{\emph{J.
  Phys. G: Nucl. Part. Phys.} (2025) 040501}.

\bibitem{em_shielding}
H.-Y.~Lin, Y.-L.~Song and L.-M.~Chang, \emph{A magnetic field cancelling system
  design for mitigating extremely low frequency magnetic field in a high tech
  fab},  in \emph{2021 Asia-Pacific International Symposium on Electromagnetic
  Compatibility (APEMC)}, pp.~1--4, 2021,
  \href{https://doi.org/10.1109/APEMC49932.2021.9596740}{DOI}.

\bibitem{6332797}
K.~Kobayashi, A.~Kon, M.~Yoshizawa and Y.~Uchikawa, \emph{Active magnetic
  shielding using symmetric magnetic field sensor method},
  \href{https://doi.org/10.1109/TMAG.2012.2197854}{\emph{IEEE Transactions on
  Magnetics} {\bfseries 48} (2012) 4554}.

\bibitem{BRYS2005527}
T.~Bryś, S.~Czekaj, M.~Daum, P.~Fierlinger, D.~George, R.~Henneck et~al.,
  \emph{Magnetic field stabilization for magnetically shielded volumes by
  external field coils},
  \href{https://doi.org/https://doi.org/10.1016/j.nima.2005.08.040}{\emph{Nuclear
  Instruments and Methods in Physics Research Section A: Accelerators,
  Spectrometers, Detectors and Associated Equipment} {\bfseries 554} (2005)
  527}.

\bibitem{Brake1991}
H.J.M.~ter Brake, H.J.~Wieringa and H.~Rogalla, \emph{Improvement of the
  performance of a mu -metal magnetically shielded room by means of active
  compensation (biomagnetic applications)},
  \href{https://doi.org/10.1088/0957-0233/2/7/004}{\emph{Meas. Sci. Technol.}
  {\bfseries 2} (1991) 596}.

\bibitem{Spemann2003}
T.~Spemann, D.and~Reinert, J.~Vogt, J.~Wassermann and B.~T., \emph{Active
  compensation of stray magnetic fields at {LIPSION}},
  \href{https://doi.org/10.1016/S0168-583X(03)01027-9}{\emph{Nucl. Instrum.
  Methods. Phys. Res. B} {\bfseries 210} (2003) 79}.

\bibitem{holmes2022}
N.~Holmes, M.~Rea, J.~Chalmers, J.~Leggett, L.J.~Edwards, P.~Nell et~al.,
  \emph{A lightweight magnetically shielded room with active shielding},
  \href{https://doi.org/https://doi.org/10.1038/s41598-022-17346-1}{\emph{Sci.
  Rep.} {\bfseries 12} (2022) 13561}.

\bibitem{shielding_overview}
Y.~Liu, J.~Yang, F.~Cao, X.~Zhang and S.~Zheng, \emph{Enhancement of magnetic
  shielding based on low-noise materials, magnetization control, and active
  compensation: A review},
  \href{https://doi.org/10.3390/ma17225469}{\emph{Materials} {\bfseries 17}
  (2024) 5469}.

\bibitem{n2EDM}
N.J.~Ayres, G.~Ban, L.~Bienstman, G.~Bison, K.~Bodek, V.~Bondar et~al.,
  \emph{The design of the {n2EDM} experiment},
  \href{https://doi.org/10.1140/epjc/s10052-021-09298-z}{\emph{Eur. Phys. J. C}
  {\bfseries 81} (2021) 512}.

\bibitem{Lauss2012}
B.~Lauss and {on behalf of the PSI UCN Project Team}, \emph{Startup of the
  high-intensity ultracold neutron source at the {Paul} {Scherrer}
  {Institute}},
  \href{https://doi.org/10.1007/s10751-012-0578-7}{\emph{Hyperfine Interact.}
  {\bfseries 211} (2012) 21}.

\bibitem{Bison2022}
G.~Bison, M.~Daum, K.~Kirch, B.~Lauss, D.~Ries, P.~Schmidt-Wellenburg et~al.,
  \emph{Ultracold neutron storage and transport at the {PSI} {UCN} source},
  \href{https://doi.org/10.1140/epja/s10050-022-00747-1}{\emph{Eur. Phys. J. A}
  {\bfseries 58} (2022) 103}.

\bibitem{Lauss2022}
B.~Lauss and B.~Blau, \emph{{UCN, the ultracold neutron source -- neutrons for
  particle physics}},  p.~004, SciPost, 2021,
  \href{https://doi.org/10.21468/SciPostPhysProc.5.004}{DOI}.

\bibitem{MSR2022}
N.J.~Ayres, G.~Ban, G.~Bison, K.~Bodek, V.~Bondar, T.~Bouillaud et~al.,
  \emph{The very large {n2EDM} magnetically shielded room with an exceptional
  performance for fundamental physics measurements},
  \href{https://doi.org/10.1063/5.0101391}{\emph{Rev. Sci. Instrum.} {\bfseries
  93} (2022) 095105}.

\bibitem{AMS2023}
C.~Abel, N.J.~Ayres, G.~Ban, G.~Bison, K.~Bodek, V.~Bondar et~al., \emph{A
  large ‘{Active} {Magnetic} {Shield}’ for a high-precision experiment},
  \href{https://doi.org/https://doi.org/10.1140/epjc/s10052-023-12225-z}{\emph{Eur.
  Phys. J. C} {\bfseries 83} (2023) 1061}.

\bibitem{1060978}
J.~Elen, W.~Franken, I.~Horvath, G.~Pasotti, M.~Ricci, J.~Roeterdink et~al.,
  \emph{The superconductor test facility sultan},
  \href{https://doi.org/10.1109/TMAG.1981.1060978}{\emph{IEEE Transactions on
  Magnetics} {\bfseries 17} (1981) 490}.

\bibitem{SCHIPPERS2007773}
J.~Schippers, R.~Dölling, J.~Duppich, G.~Goitein, M.~Jermann, A.~Mezger
  et~al., \emph{{The SC cyclotron and beam lines of PSI’s new protontherapy
  facility PROSCAN}},
  \href{https://doi.org/https://doi.org/10.1016/j.nimb.2007.04.052}{\emph{Nuclear
  Instruments and Methods in Physics Research Section B: Beam Interactions with
  Materials and Atoms} {\bfseries 261} (2007) 773}.

\bibitem{Solange2021}
S.~Emmenegger, \emph{Next Generation Active Magnetic Shielding for {n2EDM} and
  Axion-Like Particle Search}, {D}octoral {T}hesis, ETH Zurich, 2021.
\newblock https://doi.org/10.3929/ethz-b-000515206.

\bibitem{Rawlik2018}
M.~Rawlik, A.~Eggenberger, J.~Krempel, C.~Crawford, K.~Kirch, F.M.~Piegsa
  et~al., \emph{A simple method of coil design},
  \href{https://doi.org/10.1119/1.5042244}{\emph{Am. J. Phys.} {\bfseries 86}
  (2018) 602}.

\bibitem{Rawlik2018PhD}
M.~Rawlik, \emph{Active Magnetic Shielding and Axion-Dark-Matter Search},
  {D}octoral {T}hesis, ETH Zurich, 2018.
\newblock https://doi.org/10.3929/ethz-b-000273039.

\bibitem{Franke2013}
B.~Franke, \emph{Investigations of the internal and external magnetic fields of
  the neutron electric dipole moment experiment at the {Paul} {Scherrer}
  {Institute}}, {D}octoral {T}hesis, ETH Zurich, 2013.
\newblock https://doi.org/10.3929/ethz-a-010144564.

\bibitem{SFC2014}
S.~Afach, G.~Bison, K.~Bodek, F.~Burri, Z.~Chowdhuri, M.~Daum et~al.,
  \emph{Dynamic stabilization of the magnetic field surrounding the neutron
  electric dipole moment spectrometer at the {Paul} {Scherrer} {Institute}},
  \href{https://doi.org/10.1063/1.4894158}{\emph{J. Appl. Phys.} {\bfseries
  116} (2014) 084510}.

\bibitem{comsol}
{COMSOL AB}, ``{COMSOL} {Multiphysics}{\copyright}.''

\bibitem{python_mph}
N.~Stenderm, \emph{Python-{MPH}: A python interface to {COMSOL}'s {Java}
  {API}},  2023.

\bibitem{Cartan1983Geometry}
E.~Cartan and R.~Hermann, \emph{Geometry of Riemannian Spaces}, Math Sci Press,
  Massachusetts, translation or edition~ed. (1983).

\bibitem{goldberg1989}
D.E.~Goldberg and J.H.~Holland, \emph{Genetic algorithms and machine
  learning.},
  \href{https://doi.org/https://doi.org/10.1023/A:1022602019183}{\emph{Machine
  Learning 3} (1988) }.

\bibitem{hollland1975}
J.H.~Holland, \emph{Adaptation in natural and artificial systems: An
  introductory analysis with applications to biology, control, and artificial
  intelligence},
  \href{https://doi.org/https://doi.org/10.7551/mitpress/1090.001.0001}{\emph{The
  MIT Press} (1992) }.

\bibitem{haupt2004}
R.L.~Haupt and S.E.~Haupt, \emph{Practical genetic algorithms},
  \href{https://doi.org/DOI:10.1002/0471671746}{\emph{John Wiley \& Sons, Inc.}
  (2003) }.

\bibitem{liu2008}
G.~Heeralal, S.K.~Chaturvedi and A.K.~Thakur, \emph{Design and optimization of
  multilayered electromagnetic shield using a real-coded genetic algorithm},
  \href{https://doi.org/http://dx.doi.org/10.2528/PIERB12011902}{\emph{Progress
  In Electromagnetics Research B 39(39)} (2012) }.

\bibitem{Pais2018}
D.~Pais, \emph{Development of the caesium magnetometer array for the {n2EDM}
  experiment}, {D}octoral {T}hesis, ETH Zurich, 2018.
\newblock https://doi.org/10.3929/ethz-b-000511496.

\bibitem{miranda2010}
N.R.~Shaw and R.E.~Ansorge, \emph{Genetic algorithms for {MRI} magnet design},
  \href{https://doi.org/http://dx.doi.org/10.1109/TASC.2002.1018506}{\emph{IEEE
  Transactions on Applied Superconductivity} (2002) }.

\bibitem{akiba2019optuna}
T.~Akiba, S.~Sano, T.~Yanase, T.~Ohta and M.~Koyama, \emph{Optuna: A
  next-generation hyperparameter optimization framework},  in \emph{Proceedings
  of the 25th {ACM} {SIGKDD} International Conference on Knowledge Discovery \&
  Data Mining}, pp.~2623--2631, ACM, 2019,
  \href{https://doi.org/10.1145/3292500.3330701}{DOI}.

\end{thebibliography}\endgroup

\end{document}